\documentclass[12pt]{article}
\usepackage[pdftex]{graphicx}
\begin{document}
\begin{center}

{\bf Singly differential cross sections with exchange for Ps-fragmentation }\\

\vskip 0.5cm
{\bf Hasi Ray}\footnote[1]{Present address: Gravitation Group, TIFR, Mumbai, India} \\

S-407, Flat No. 6, B. P. Township, Kolkata 700094, India

Email: hasi\_ray@yahoo.com

\end{center}
\vskip 1.0cm

{\bf Abstract}: 
 Ps ionization in Ps-atom scattering is of fundamental importance. The singly differential cross sections (SDCS) provides more accurate information to test a theory than integrated or total ionization cross section since the averaging over one parameter is not required. We evaluate the SDCS for Ps-ionization with respect to the longitudinal energy distribution of the break-up positron and electron in Ps-H and Ps-He scattering and compare them with the recently available experimental and theoretical data. 
\vskip 0.5cm
{\bf PACS No.} 36.10.Dr, 34.70.+e, 82.30.Gg \\

{\bf Keywords:} Positronium (Ps), ionization, differential cross section, Coulomb-Born-Oppenheimer approximation, static Coulomb interaction.
\vskip 0.5cm

Positronium (Ps) is the lightest hydrogen-like exotic atom. 
 It is itself its anti-atom. The charge and mass centers of it coincide. The interesting
property is that the mean static Coulomb interaction vanishes when Ps interacts with an atom / molecule. Its polarizability is eight times higher than that of H. Two kinds of it
are known as para-Ps and ortho-Ps depending on spin states of positron, $e^+$ and electron, $e^-$.
The fragmentation of positronium (Ps) is a process which helps to understand the mechanism of Ps and atom/molecule scattering [1-13]. Recently, the experimental group at University College of London (UCL) lead by G. Laricchia has measured [4-6] the singly-differential Ps-ionization cross sections with respect to the longitudinal energy distribution of the break-up positron in Ps and He collision which motivated the present study. The number of papers which deal Ps-fragmentation is extremely limited [2-20]. Only a very few experimental data [4-6] on Ps-ionization in Ps-He scattering is available in the literature. The fragmentation or ionization or break-up of Ps starts only at 6.8 eV. \\

 We study collision of Ps with atom. The kind of calculation is a bit difficult and challenging due to presence of many charge centers. In such a system of a Ps and an atom, the exchange between the Ps-electron and the atomic-electron is highly important and sensitive at lower incident energies, and it is an effect of fundamental interest.
We calculate the singly (energy) differential cross sections for Ps-ionization in Ps-H and Ps-He scattering using the Coulomb-Born-Oppenheimer approximation (CBOA) and the Coulomb-Born approximation (CBA); both the theories were introduced by Ray [2,3,15-19] to calculate the integrated cross sections.
The differential cross section is more informative than integrated cross section since it relaxes the averaging over a coordinate system e.g. angular or energy distribution. In the present case the averaging over energy distribution of the break-up positron and/or electron is relieved.\\

The scattering amplitudes can be expressed in post and prior forms, both will provide the same results if the system wavefunctions are exact [21]. We use the post form of the scattering amplitude; in this expression $\langle \Psi_f\mid v_f\mid\Psi_i^+\rangle$ (following conventional notation), the incident channel wave function is treated as a plane wave as in first Born approximation (FBA). Since the projectile is a neutral system, the present plane wave approximation for the incident Ps may not introduce an error. In addition the first order effect due to polarizability is absent in such system. So it is nearly accurate which is evident from the comparison of FBA and the close coupling approximation (CCA) results of our earlier calculation [22,23]. Only at very low energies, the FBA and CCA elastic cross sections differ in Ps-H [22] and Ps-Li [23] systems. The break-up electron is treated as a Coulomb wave. As the incident Ps is a composite system of a positron and an electron, we treat the moving Ps as having a center of mass (c.m.) motion with momenta $\bf k_i$ and $\bf k_f$ in the initial and final channels respectively, and a relative motion of electron with respect to positron; the c.m. motion is represented by a plane wave. \\

A few words regarding the target He wave function is useful. The accuracy of the present calculation again depends on the accuracy of the ground state He atomic wavefunction. We used the first kind of wavefunction prescribed by Winter [24]. This wave function is not correlated. A correlated wave function for the target atomic He would be more useful for better accuracy, but it complicates our calculation further. It should be a good task to use a correlated wave function to tackle the present problem.\\

 The singly energy differential cross section (SDCS) for the break-up of Ps in Ps and atom scattering is defined [10] as 
$$\frac{d\sigma}{dE} = \int d\hat{\bf k}_p \int d\hat{\bf k}_e
     \frac{d^3\sigma}{d\hat{\bf k_p}d\hat{\bf k}_e dE} \eqno(1)$$ \\
where ${\bf k}_p$, ${\bf k}_e$ are the momenta of the break-up positron and electron respectively and E is the sum of the kinetic energies of break-up positron and electron after ionization. \\

In inelastic collisions, only the energy conservation is true. If $v_p$, $v_e$ represent respectively the speeds (i.e. magnitudes of the velocities) of break-up positron and electron from Ps, $\epsilon_{Ps}$ is the ionization potential of Ps, then the energy conservation relation
$$ E = E_i - \epsilon_{Ps} = \frac{1}{2}v_p^2 + \frac{1}{2}v_e^2   \eqno(2a) $$
             $$ = v_f^2 + \frac{1}{4} v^2      \eqno(2b) $$\\
should be fulfilled if the incident energy is $E_i$; $v_f$ is the magnitude of the c.m. velocity of fragmented Ps and $v$ is the magnitude of relative velocity of the break-up electron with respect to the break-up positron in the final channel after ionization. \\

  The momentum transfer ${\bf Q}$ is defined as ${\bf Q} = {\bf k}_i -{\bf k}_f$, so that 
$$ Q^2 = k_i^2 + k_f^2 - 2k_ik_fcos{\theta} \eqno(3) $$ 
if $\theta$ is the scattering angle and ${\bf k}_i$, ${\bf k}_f$ are initial and final momenta. According to our definition, the SDCS is defined as

         $$ \frac{d\sigma}{dE_{\bf k}} = \int d\hat{\bf k}_f \int d\hat{\bf k}
     \frac{d^3\sigma}{d\hat{\bf k_f}d\hat{\bf k} dE_{\bf k}}    \eqno(4a)$$
as $$ TDCS = \frac{d^3\sigma}{d\hat{\bf k_f}d\hat{\bf k} dE_{\bf k}}  \eqno(4b)$$
$$ = \frac{k_f k}{k_i} \mid \{Amplitude\}_{Scattering} \mid ^2 \eqno(4c) $$ 
Again 
    $$ {\bf k}_p = {\bf k}_f - \frac{1}{2}{\bf k} \eqno(5a)$$
    $$ {\bf k}_e = {\bf k}_f + \frac{1}{2}{\bf k} \eqno(5b)$$ \\
which give $d{{\bf k}_p} = d{{\bf k}_f}$ if we make ${\bf k}$ unaltered and $d{{\bf k}_e} = \frac{1}{2} d{{\bf k}}$ if we make ${\bf k}_f$ unaltered. It is obvious that $ dE_{\bf k} = k dk$ and $\int dE_k = \int dE$. \\
    
If the longitudinal energy of break-up positron is $E_{pl}$ and it makes an angle $\theta_p$ with ${\bf k_i}$, then
$$ E_p = \frac{1}{2}v_p^2 = \frac{E_{pl}}{cos^2{\theta_p}}  \eqno(6a) $$

and
$$ v_e^2 = 2(E_i - \epsilon_{Ps} - \frac{E_{pl}}{cos^2{\theta_p}})  \eqno(6b) $$ \\

The magnitude of relative velocity (i.e. relative speed) of break-up electron with respect to break-up positron can be defined as, 
             $$ v = v_e - v_p  \eqno(7a)$$
 $$ v^2 = v_p^2 + v_e^2 - 2v_pv_e = 4k^2  \eqno(7b) $$\\
and 
$$v_f^2 = E_i - \epsilon_{Ps} - \frac{1}{4}v^2 = \frac{1}{4}k_f^2\eqno(7c)$$\\

If the longitudinal energy of break-up electron is $E_{el}$ and it makes an angle $\theta_e$ with ${\bf k_i}$, then
$$ E_e = \frac{1}{2}v_e^2 = \frac{E_{el}}{cos^2{\theta_e}}  \eqno(8a) $$\\
and
$$ v_p^2 = 2(E_i - \epsilon_{Ps} - \frac{E_{el}}{cos^2{\theta_e}})\eqno(8b)$$\\

So the singly differential cross sections with respect to the longitudinal energies of the break-up positron and the break-up electron should be defined as
$$\frac{d\sigma}{dE_{pl}} = \frac{1}{2} \int d\hat{\bf k}_f \int d\hat{\bf k} (\frac{ TDCS}{cos^2\theta_p}) \eqno(9) $$
$$\frac{d\sigma}{dE_{el}} = \frac{1}{2} \int d\hat{\bf k}_f \int d\hat{\bf k}
(\frac{ TDCS}{cos^2\theta_e}) \eqno(10) $$ \\

 The triply differential cross sections (TDCS) for the break-up of Ps in Ps-H and Ps-He scatterings are defined as
$$ \frac{d^3\sigma}{d\hat{\bf k_f}d\hat{\bf k}dE_{\bf k}} = \frac{k_f k}{k_i}\{\frac{1}{4}\mid F_k + G_k \mid^2 + \frac{3}{4}\mid F_k - G_k \mid^2 \} \eqno(11a) $$
$$ \frac{d^3\sigma}{d\hat{\bf k_f}d\hat{\bf k}dE_{\bf k}} = \frac{k_f k}{k_i}\mid F_{\bf k}^{He} - G_{\bf k}^{He} \mid^2  \eqno(11b) $$ 
\\ 
where $F_k$, $G_k$ represent respectively the direct and exchange matrix elements for Ps-H scattering; $F_{\bf k}^{He}$, $G_{\bf k}^{He}$ represent respectively the direct and exchange matrix elements for Ps-He scattering. These are defined as
 
$$F_{\bf k}^{He}(\hat{\bf k}_f) = -\frac{1}{\pi}\int e^{-i{\bf k_f.R_1}}\eta_{\bf k}^*(\mbox{\boldmath$\rho$}_1)\Phi_f^*\{{\bf r_2, r_3}\}[V_{He}^F]e^{i{\bf k_i.R_1}}\eta_{1s}(\mbox{\boldmath$\rho$}_1)\Phi_i\{{\bf r_2, r_3}\}d{\bf x}d{\bf r_1}d{\bf r_2}d{\bf r_3} \eqno(12a) $$
$$G_{\bf k}^{He}(\hat{\bf k}_f) = -\frac{1}{\pi}\int e^{-i{\bf k_f.R_2}}\eta_{\bf k}^*(\mbox{\boldmath$\rho$}_2)\Phi_f^*\{{\bf r_1, r_3}\}[V_{He}^G]e^{i{\bf k_i.R_1}}\eta_{1s}(\mbox{\boldmath$\rho$}_1)\Phi_i\{{\bf r_2, r_3}\}d{\bf x}d{\bf r_1}d{\bf r_2}d{\bf r_3} \eqno(12b) $$
with
$$ V_{He}^F = \frac{Z}{\mid{\bf x}\mid}-\frac{Z}{\mid{\bf r_1}\mid}-\frac{1}{\mid{{\bf x}-{\bf r_2}}\mid}+\frac{1}{\mid{{\bf r_1}-{\bf r_2}}\mid}-\frac{1}{\mid{{\bf x}-{\bf r_3}}\mid}+\frac{1}{\mid{{\bf r_1}-{\bf r_3}}\mid} \eqno(13a) $$
and
$$ V_{He}^G = \frac{Z}{\mid{\bf x}\mid}-\frac{Z}{\mid{\bf r_2}\mid}-\frac{1}{\mid{{\bf x}-{\bf r_1}}\mid}+\frac{1}{\mid{{\bf r_2}-{\bf r_1}}\mid}-\frac{1}{\mid{{\bf x}-{\bf r_3}}\mid}+\frac{1}{\mid{{\bf r_2}-{\bf r_3}}\mid} \eqno(13b) $$
\\
with
   ${\bf R}_j = \frac{1}{2}({\bf x}+{\bf r}_j)$   
     and 
$\mbox{\boldmath$\rho$}_j = ({\bf x} - {\bf r}_j)$; j=1,2. Here,
${\bf x}$ is the coordinate of positron in Ps, and 
${\bf r}_j$; j = 1 to 3 are those of electrons in Ps and He respectively in the incident channel with respect to the center of mass of the system. Functions 
$\eta$ and $\Phi$ indicate the wave functions of Ps and He respectively. Subscript `$i$` identifies the incident channel, whereas `$f$` represents the final channel. Accordingly ${\bf k}_i$ and ${\bf k}_f$ are the momenta of the projectile in the initial and final channels respectively. Z is the nuclear charge of the target helium atom. The target atomic wave functions are considered at ground states in both the incident and final channels. If we remove the third electron from the Ps-He system which is represented by the position coordinate ${\bf r}_3$, the expressions (12a), (12b), (13a), (13b) should fit to Ps-H system. The continuum Coulomb wave function $\eta_{\bf k}$ is chosen from Ref.[3] which is orthogonal to the ground state Ps wave function.\\

 In figures 1-5, we report the present SDCS results for Ps-H scattering with respect to the longitudinal energy distribution of the break-up positron and electron at the incident energies of Ps e.g. 13 eV, 18 eV, 25 eV, 33 eV, 60 eV. For the Ps-H system, there is no experimental data and no reported theoretical data. The figures 6-11 display the present SDCS results for Ps-He system with respect to the longitudinal energy distribution of break-up positron and electron respectively with the recently available corresponding experimental [5,6] and theoretical [10,11,24] data at different incident energies. It is to be noted that the area under the SDCS curves for $e^+$ and $e^-$ are equal. So we can get the same values of total cross section $\sigma$ using both the SDCSs and it indicates that we formulate the SDCSs correctly.\\

 The present singly differential cross sections for Ps-He system at different incident Ps energies with respect to the longitudinal energy distribution of break-up positron are compared with the available experimental and the impulse approximation (IA) data. No experimental data are available for the singly differential cross section with respect to the longitudinal energy distribution of break-up electron for Ps-He system; these are compared only with the recently available IA data of Walters et al. [10,11]. However, at the incident enegy of 60 eV, we have only the experimental data for $e^+$-distribution and no other theoretical data for both $e^+$ and $e^-$ distributions; the present results are compared with experimental data and displayed in figure 10. In figure 11, we presented the data at the incident energy = 100 eV; we compare our results with the theoretical data of Starrett et al [20]. There are no experimental data at 100 eV.\\

From the the comparison of our SDCS data using CBA and CBOA theories for the longitudinal energy distributions of the break-up positron and the break-up electron, one can get a very good idea about the effect of exchange on ionization of Ps in Ps and atom scattering. The exchange effects are stronger at lower incident energies and at the lower energies of the break-up positron and electron. The agreement of the present CBOA data for Ps-He system with the experiment is improving gradually with the increase of the incident energy of the incoming Ps from 13 eV to 60 eV and with the increase of the longitudinal energies of the break-up positron from very close to zero to $E = E_i - \epsilon_{Ps}$. Both the findings are consistent with the existing physics. Our CBA data seem to be more closer to the IA data than our CBOA data.\\
  
In conclusion we have successfully evaluated the SDCSs for Ps-ionization in Ps-H and Ps-He scattering at different incident Ps beam energies using the CBOA and CBA theories for both the positron and electron energy distributions to compare them with the existing experimental [5,6] and the theoretical [20] data. Our theoretical data are showing better agreement with the experimental data at higher incident energies of the projectile and higher longitudinal energies of the break-up positron and are consistent with the existing physics. The use of a correlated wavefunction for the target helium atom and the coupling effect among various channels are expected to improve the results at lower energies; however that will complicate the calculation.\\

Author is thankful to C. S. Unnikrishnan for providing the scope to visit his laboratory at TIFR, Mumbai for three weeks in August 2008 and the facilities to revise the manuscript.\\
 
{\bf REFERENCES:}

[1] H.S.W.Massey and C.B.O.Mohr, Proc. Phys. Soc. {\bf 67} 695 (1954).

[2] H.Ray, Euro. Phys. Lett. {\bf 73} 21 (2006).

[3] H.Ray, PRAMANA {\bf 66} 415 (2006).

[4] S.Armitage, D.E.Leslie, A.J.Garner and G. Laricchia, Phys. Rev. Lett {\bf 89} 173402 (2002).

[5] S. Amritage, D. E. Leslie, J. Beale, G. Laricchia, NIMB, {\bf 247} 98 (2006). 

[6] G. Laricchia at UCL, London (2005-2006) private communication.

[7] C.P.Campbell, M.T.McAlinden, F.G.R.S.MacDonald and H.R.J.Walters, Phys. Rev. Lett. {\bf 80} 5097 (1998). 

[8] J.E.Blackwood, M.T.McAlinden and H.R.J.Walters, Phys. Rev. A {\bf 65}, 032517 (2002).

[9] J. E. Blackwood, C.P.Campbell, M.T.McAlinden and H.R.J.Walters, Phys. Rev. A {\bf 60} 4454 (1999). 

[10] C. Starrett, Mary T. McAlinden and H. R. J. Walters, Phys. Rev. A. {\bf 72} 012508 (2005). 

[11] H.R.J.Walters, C.Starrett and M.T.McAlinden, NIMB, {\bf 247} 111 (2006). 

[12] L. Sarkadi, Phys. Rev. A {\bf 68} 032706 (2003).

[13] P. K. Biswas and Sadhan K. Adhikari, Phys. Rev. A {\bf 59} 363 (1999).

[14] M.T.McAlinden, F.G.R.S.MacDonald and H.R.J.Walters, Can. J. Phys. {\bf 74} 434 (1996). 

[15] H.Ray, PRAMANA {\bf 63} 1063 (2004).

[16] H.Ray, J.Phys.B {\bf 35} 3365 (2002).

[17] H.Ray, NIMB {\bf 192} 191 (2002).

[18] H.Ray, Phys. Lett. A {\bf 299} 65 (2002).

[19] H.Ray, Phys. Lett.A {\bf 252} 316 (1999).

[20] C. Starrett and H.R.J.Walters, J. Elec. Spec. {\bf 161} 194 (2007).

[21] D.P.Dewangan and J.Eichler, Phys. Rep. {\bf 247} 59 (1994).

[22] H.Ray and A.S.Ghosh, J. Phys. B. {\bf 31} 4427 (1998).

[23] H. Ray, J. Phys. B {\bf 33} 4285 (2000).

[24] T. G. Winter and C. C. Lin, Phys. Rev. A {\bf 12} 434 (1975). \\

\pagebreak
{\bf Figure Captions:}\\

{\bf Figure 1:}
Singly differential Ps-ionization cross sections in Ps-H scattering with respect to the longitudinal energies of the break-up positron: (a) the solid curve using CBOA, (b) the dashed and dotted curve using CBA. Similarly SDCS distributions for the break-up electron: (c) the dotted curve using CBOA, (d) the dashed and small dashed curve using CBA, at the incident energy 13 eV.\\

{\bf Figure 2:}
Singly differential Ps-ionization cross sections in Ps-H scattering with respect to the longitudinal energies of the break-up positron: (a) the solid curve using CBOA, (b) the dashed and dotted curve using CBA. Similarly SDCS distributions for the break-up electron: (c) the dotted curve using CBOA, (d) the dashed and small dashed curve using CBA, at the incident energy 18 eV.\\

{\bf Figure 3:}
Singly differential Ps-ionization cross sections in Ps-H scattering with respect to the longitudinal energies of the break-up positron: (a) the solid curve using CBOA, (b) the dashed and dotted curve using CBA. Similarly SDCS distributions for the break-up electron: (c) the dotted curve using CBOA, (d) the dashed and small dashed curve using CBA, at the incident energy 25 eV.\\

{\bf Figure 4:}
Singly differential Ps-ionization cross sections in Ps-H scattering with respect to the longitudinal energies of the break-up positron: (a) the solid curve using CBOA, (b) the dashed and dotted curve using CBA. Similarly SDCS distributions for the break-up electron: (c) the dotted curve using CBOA, (d) the dashed and small dashed curve using CBA, at the incident energy 33 eV.\\

{\bf Figure 5:}
Singly differential Ps-ionization cross sections in Ps-H scattering with respect to the longitudinal energies of the break-up positron: (a) the solid curve using CBOA, (b) the dashed and dotted curve using CBA. Similarly SDCS distributions for the break-up electron: (c) the dotted curve using CBOA, (d) the dashed and small dashed curve using CBA, at the incident energy 60 eV.\\

{\bf Figure 6:}
Singly differential Ps-ionization cross sections in Ps-He scattering with respect to the longitudinal energies of break-up positron and electron at the incident energy 13 eV. The solid curve is the present CBOA results for positron and the dotted one for electron. The long-dashed curve is the IA-results of Belfast group [10] for positron and short-dashed one are the same for electron. The solid squares are the experimental points [6] with error bars. The dashed dotted and dashed small dashed are the CBA data for positron and electron.\\

{\bf Figure 7:}
Singly differential Ps-ionization cross sections in Ps-He scattering with respect to the longitudinal energies of break-up positron and electron at the incident energy 18 eV.  The solid curve is the present CBOA results for positron and the dotted one for electron. The long-dashed curve is the IA-results of Belfast group [10] for positron and short-dashed one are the same for electron. The solid squares are the experimental points [6] with error bars. The dashed dotted and dashed small dashed are the CBA data for positron and electron.\\

{\bf Figure 8:}
Singly differential Ps-ionization cross sections in Ps-He scattering with respect to the longitudinal energies of break-up positron and electron at the incident energy 25 eV.  The solid curve is the present CBOA results for positron and the dotted one for electron. The long-dashed curve is the IA-results of Belfast group [10] for positron and short-dashed one are the same for electron. The solid squares are the experimental points [6] with error bars. The dashed dotted and dashed small dashed are the CBA data for positron and electron.\\

{\bf Figure 9:} 
Singly differential Ps-ionization cross sections in Ps-He scattering with respect to the longitudinal energies of break-up positron and electron at the incident energy 33 eV.  The solid curve is the present CBOA results for positron and the dotted one for electron. The long-dashed curve is the IA-results of Belfast group [10] for positron and short-dashed one are the same for electron. The solid squares are the experimental points [6] with error bars. The dashed dotted and dashed small dashed are the CBA data for positron and electron.\\

{\bf Figure 10:} 
Singly differential Ps-ionization cross sections in Ps-He scattering with respect to the longitudinal energies of break-up positron and electron at the incident energy 60 eV.  The solid curve is the present CBOA results for positron and the dotted one for electron. The solid squares are the experimental points [6] with error bars. The dashed dotted and dashed small dashed are the CBA data for positron and electron.\\

{\bf Figure 11:} Singly differential Ps-ionization cross sections in Ps-He scattering with respect to the longitudinal energies of break-up positron and electron at the incident energy 100 eV.  The solid curve is the present CBOA results for positron and the dotted one for electron. The long-dashed curve is the IA-results of Belfast group [20] for positron and short-dashed one are the same for electron. The dashed dotted and dashed small dashed are the CBA data for positron and electron.\\

\begin{figure}
\includegraphics[width=17.0cm,height=19.0cm]{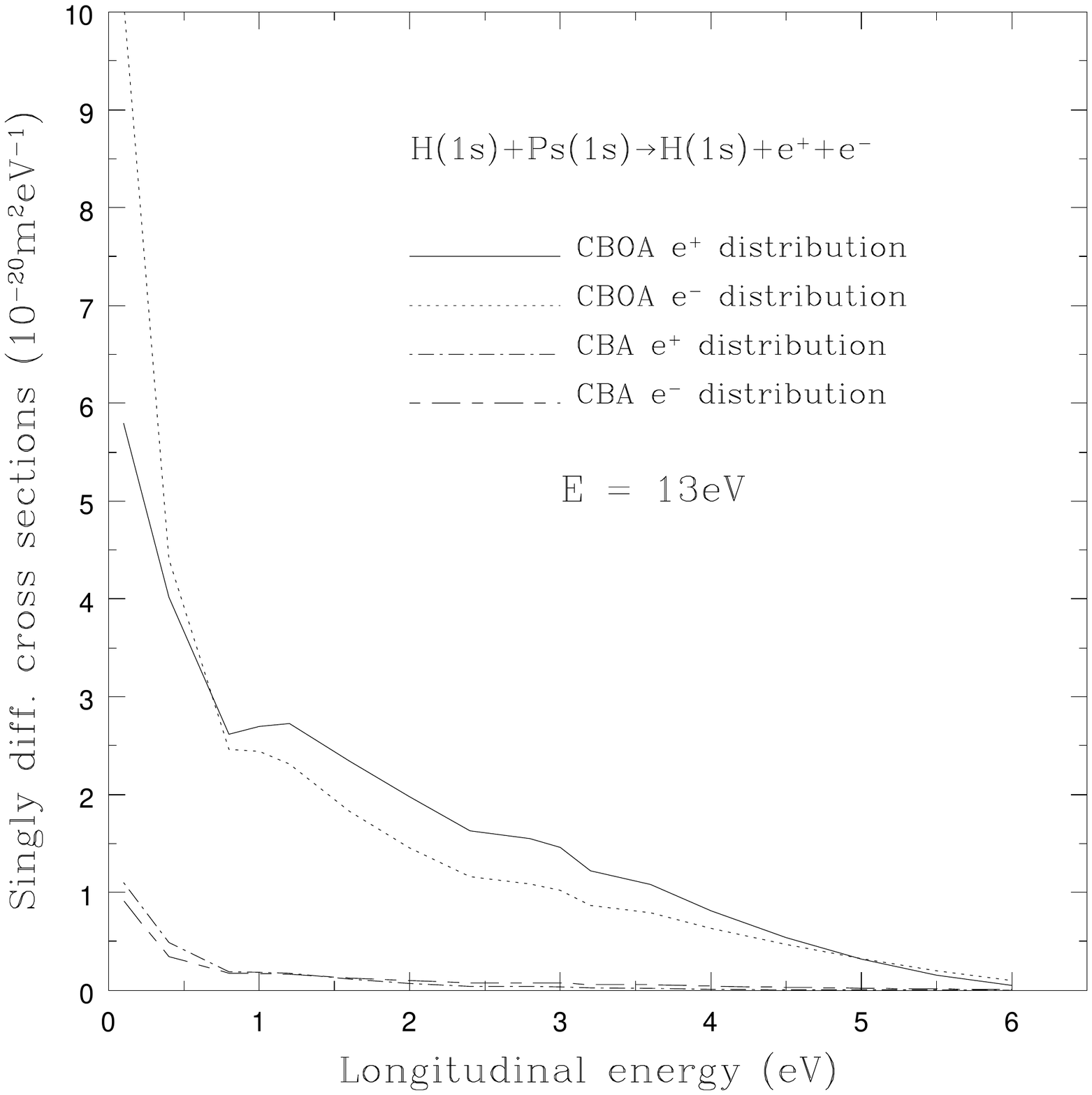}
\caption{}
\end{figure}
\begin{figure}
\includegraphics[width=17.0cm,height=19.0cm]{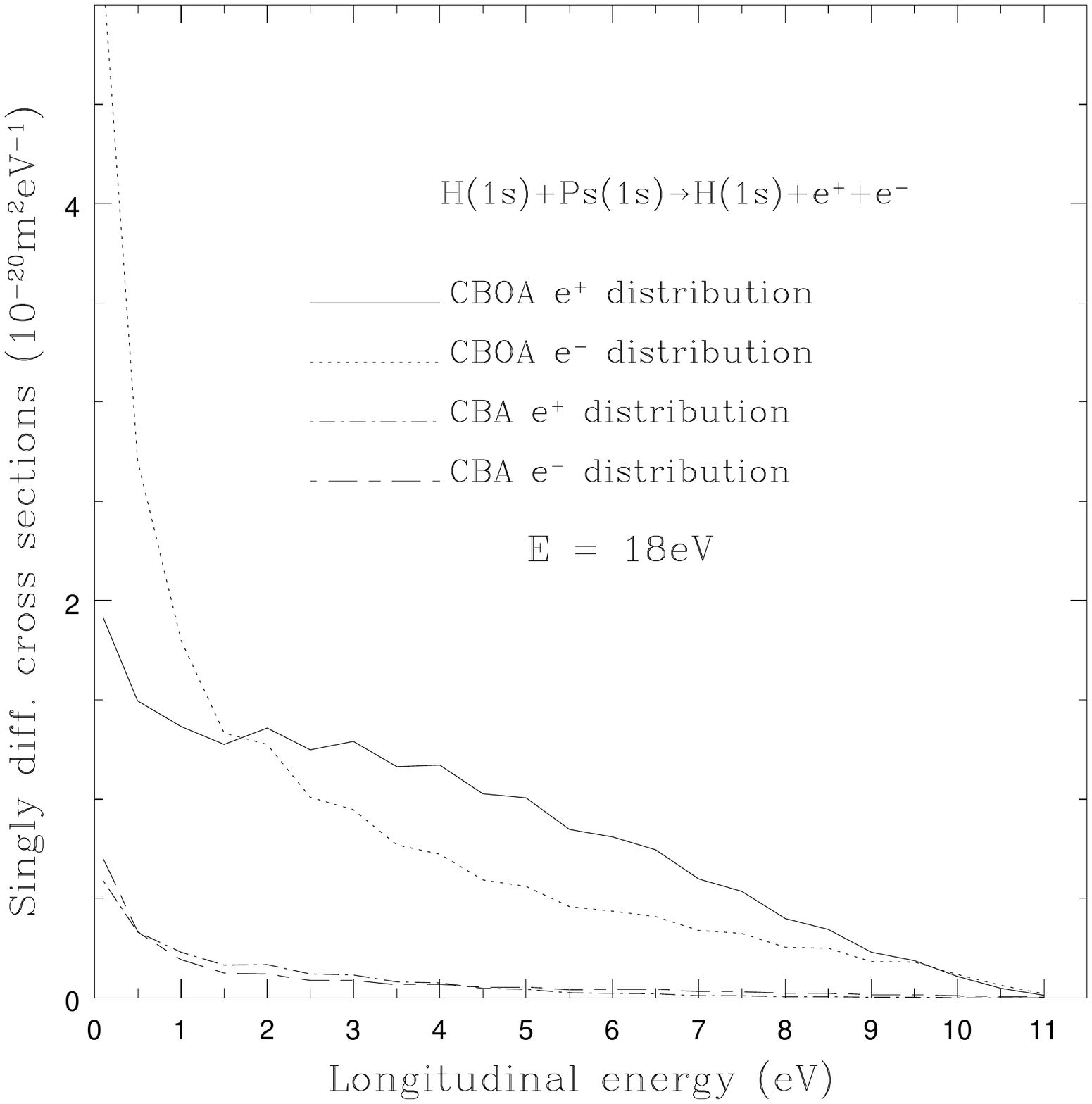}
\caption{}
\end{figure}
\begin{figure}
\includegraphics[width=17.0cm,height=19.0cm]{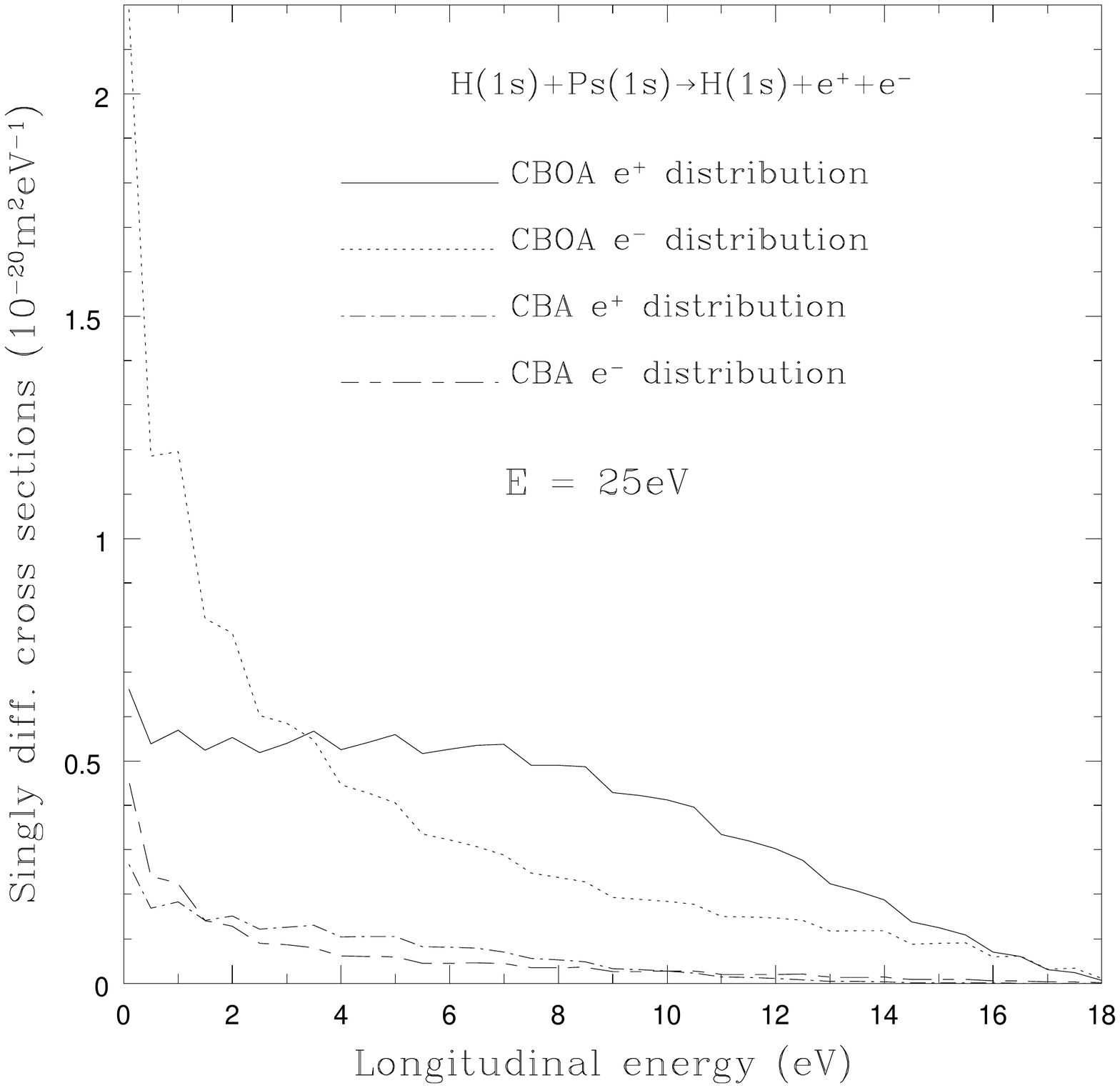}
\caption{}
\end{figure}
\begin{figure}
\includegraphics[width=17.0cm,height=19.0cm]{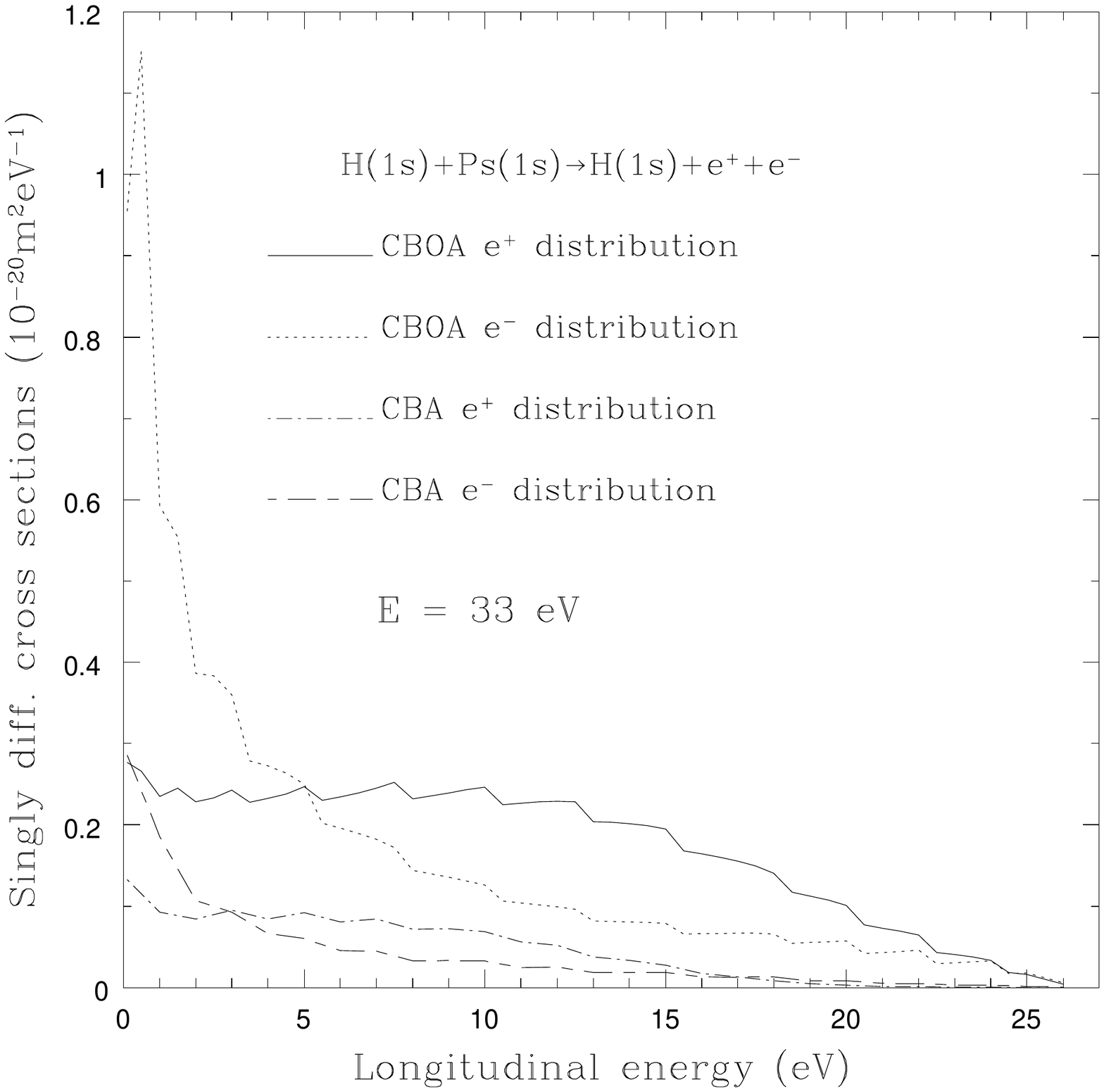}
\caption{}
\end{figure}
\begin{figure}
\includegraphics[width=17.0cm,height=19.0cm]{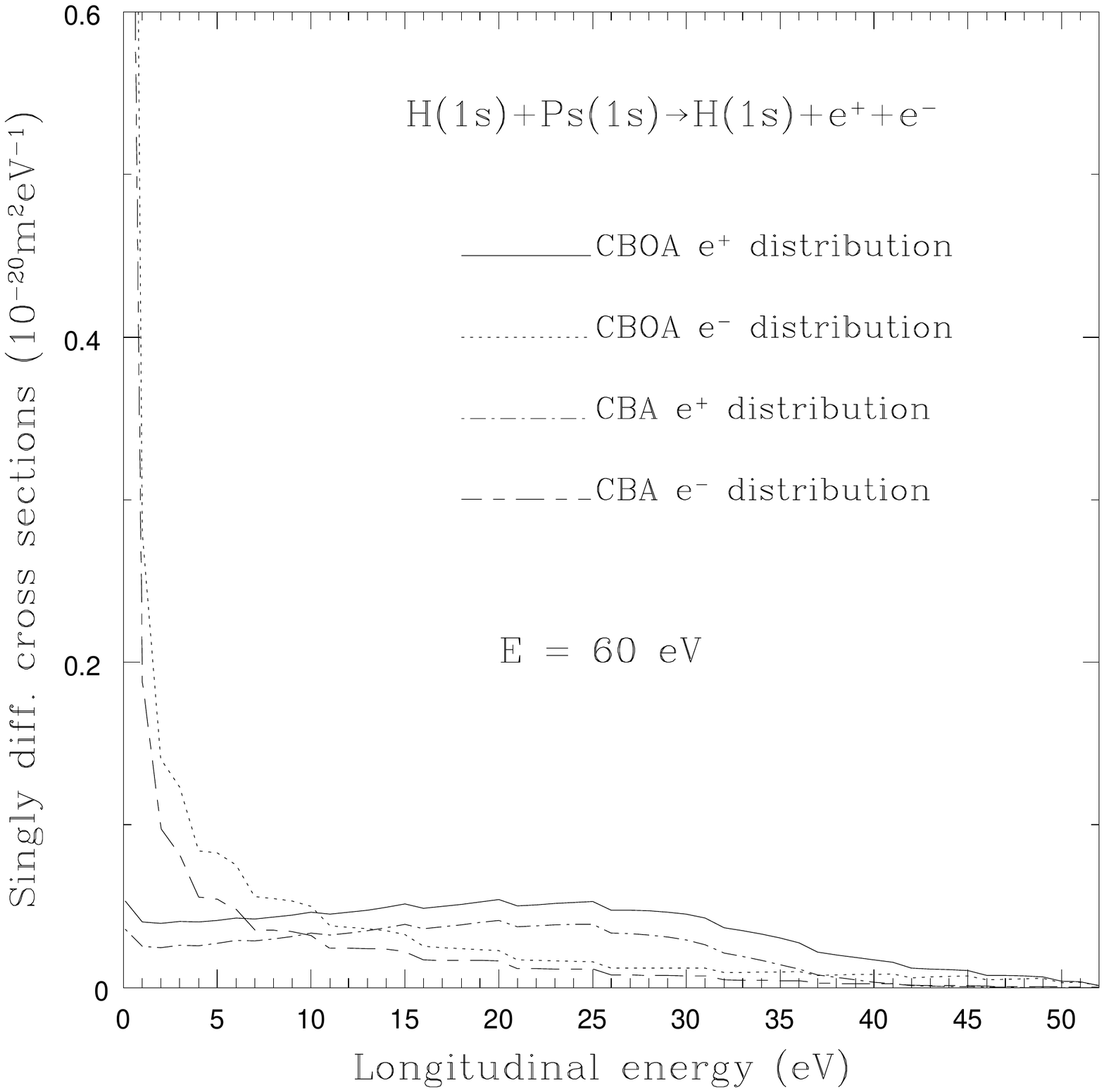}
\caption{}
\end{figure}
\begin{figure}
\includegraphics[width=17.0cm,height=19.0cm]{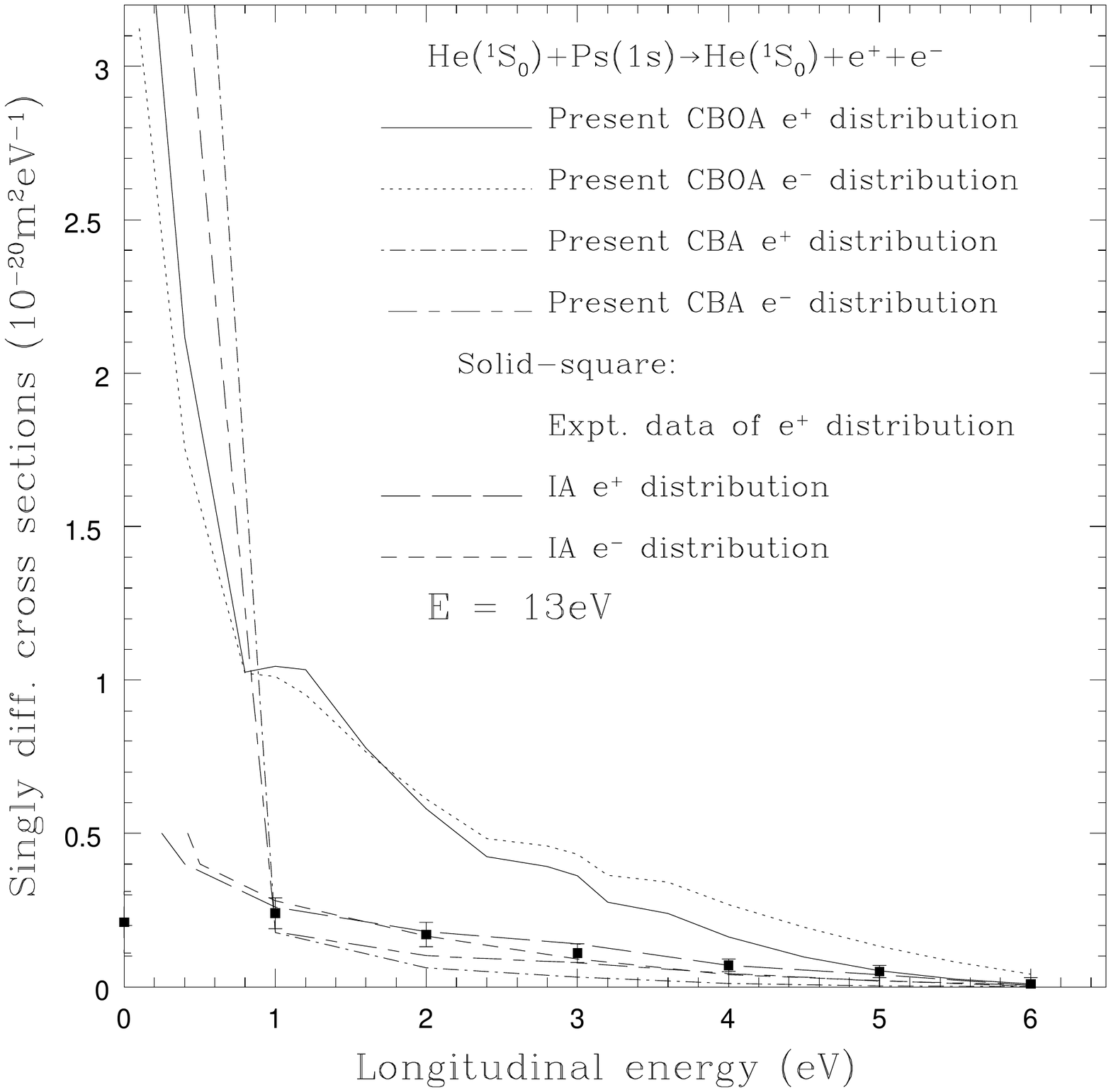}
\caption{}
\end{figure}
\begin{figure}
\includegraphics[width=17.0cm,height=19.0cm]{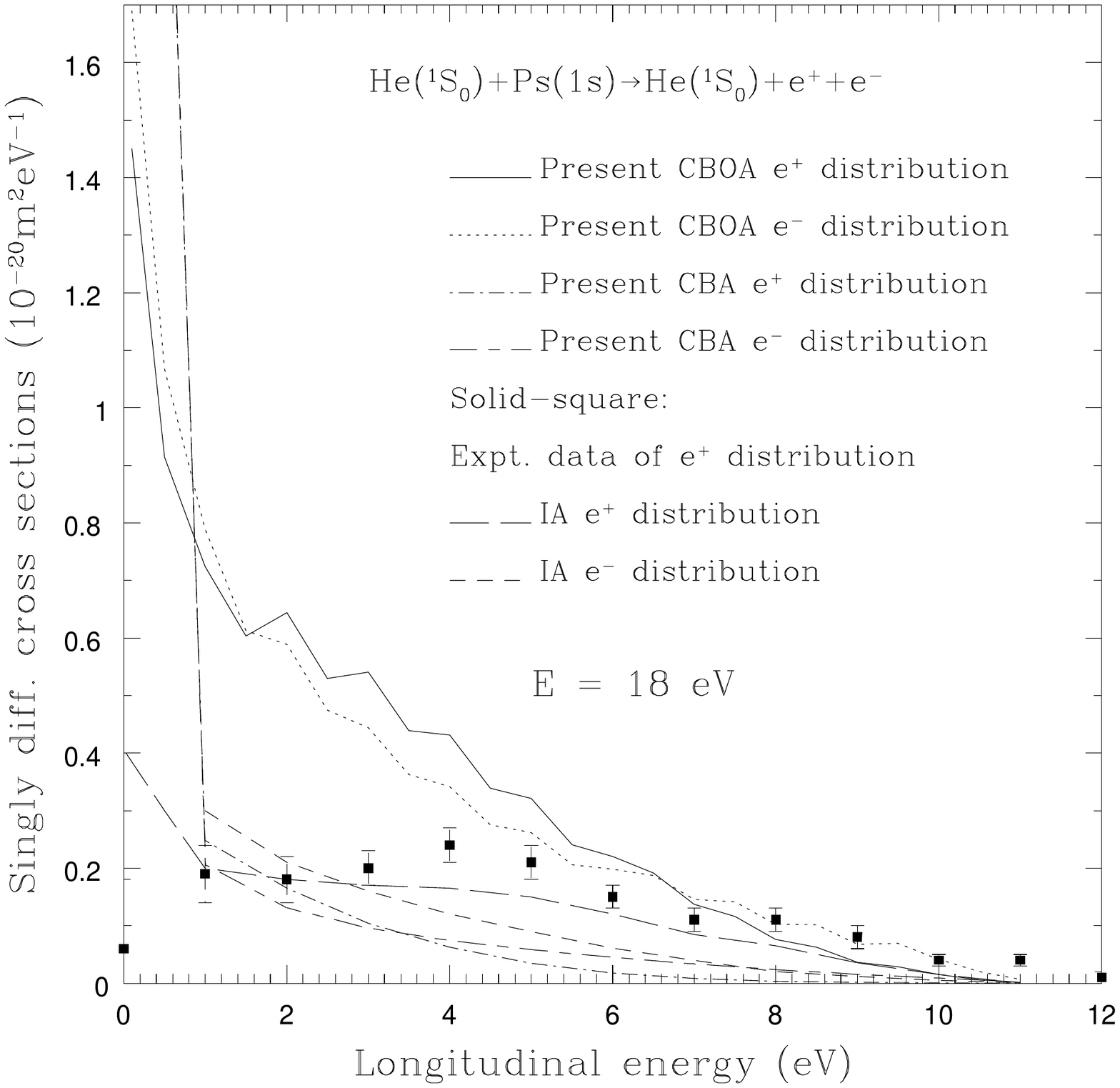}
\caption{}
\end{figure}
\begin{figure}
\includegraphics[width=17.0cm,height=19.0cm]{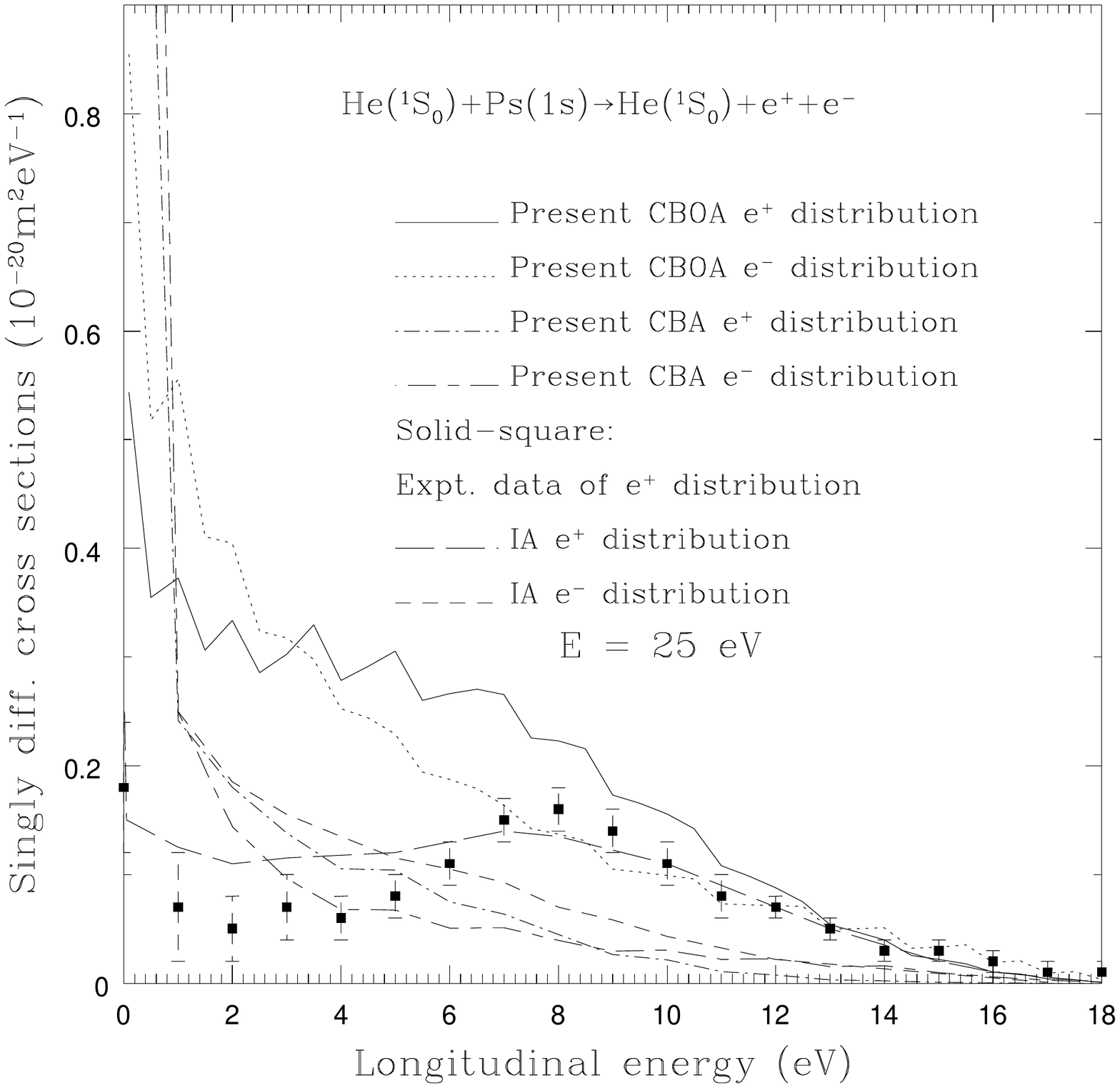}
\caption{}
\end{figure}
\begin{figure}
\includegraphics[width=17.0cm,height=19.0cm]{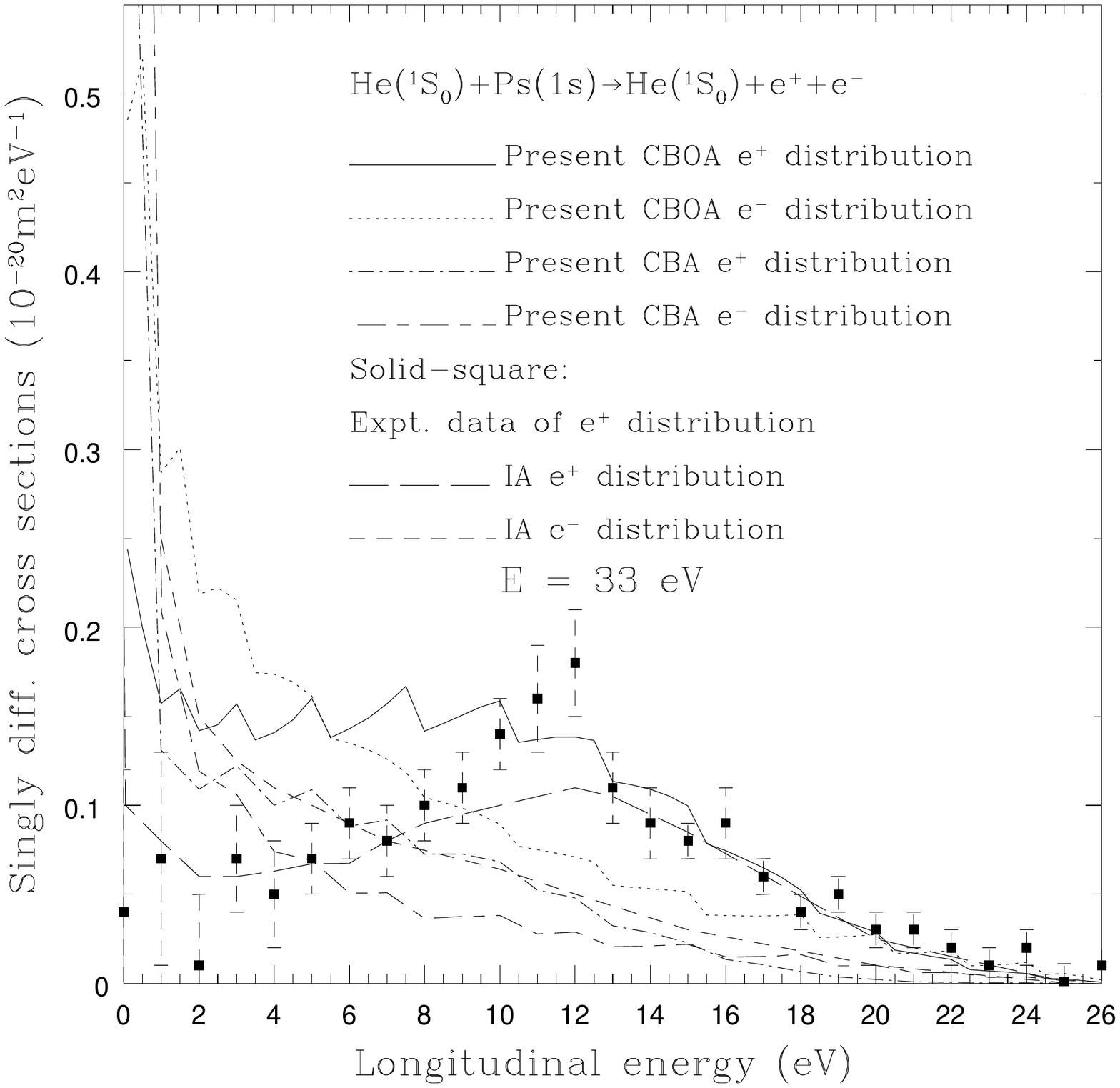}
\caption{}
\end{figure}
\begin{figure}
\includegraphics[width=17.0cm,height=19.0cm]{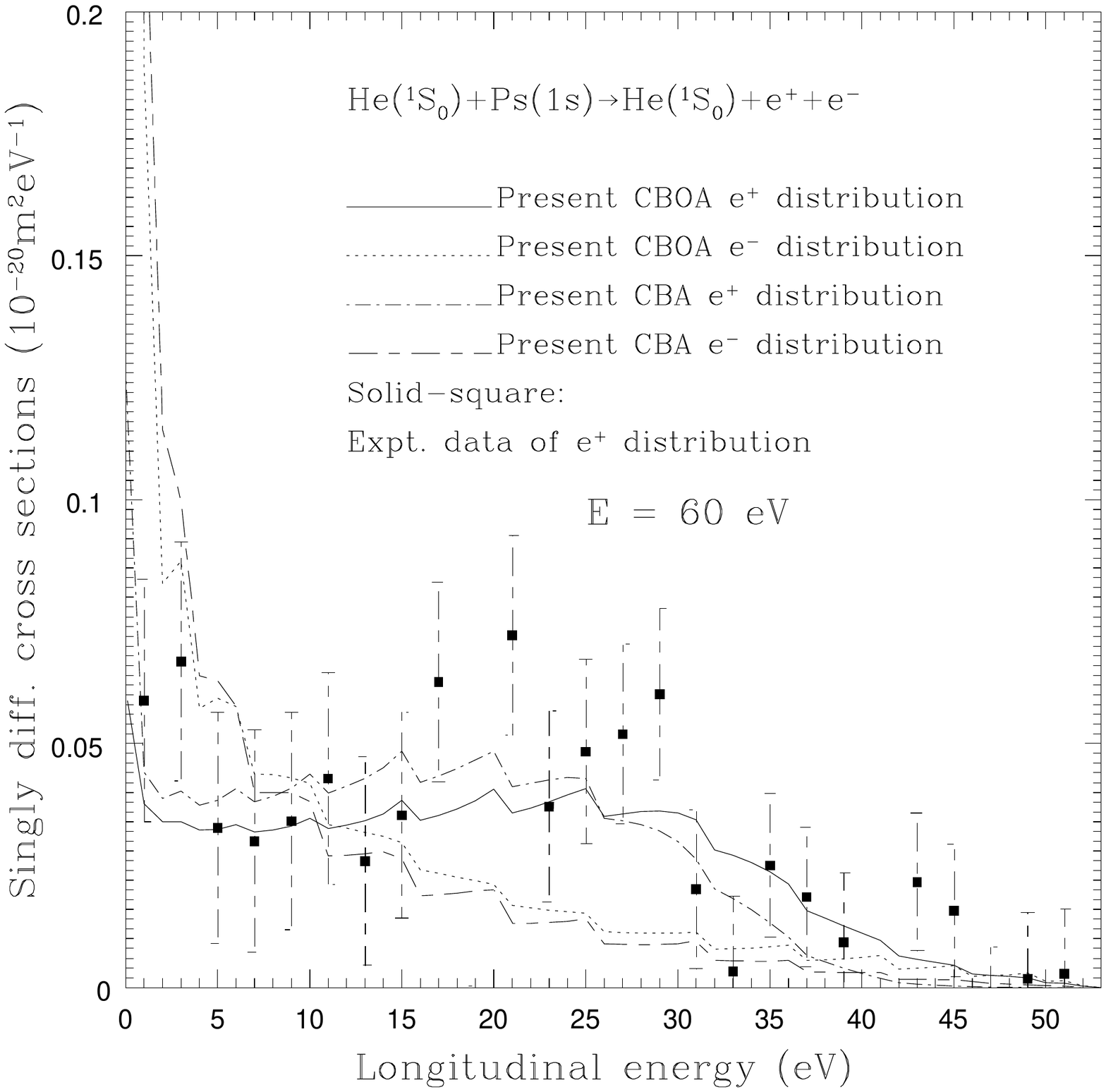}
\caption{}
\end{figure}
\begin{figure}
\includegraphics[width=17.0cm,height=19.0cm]{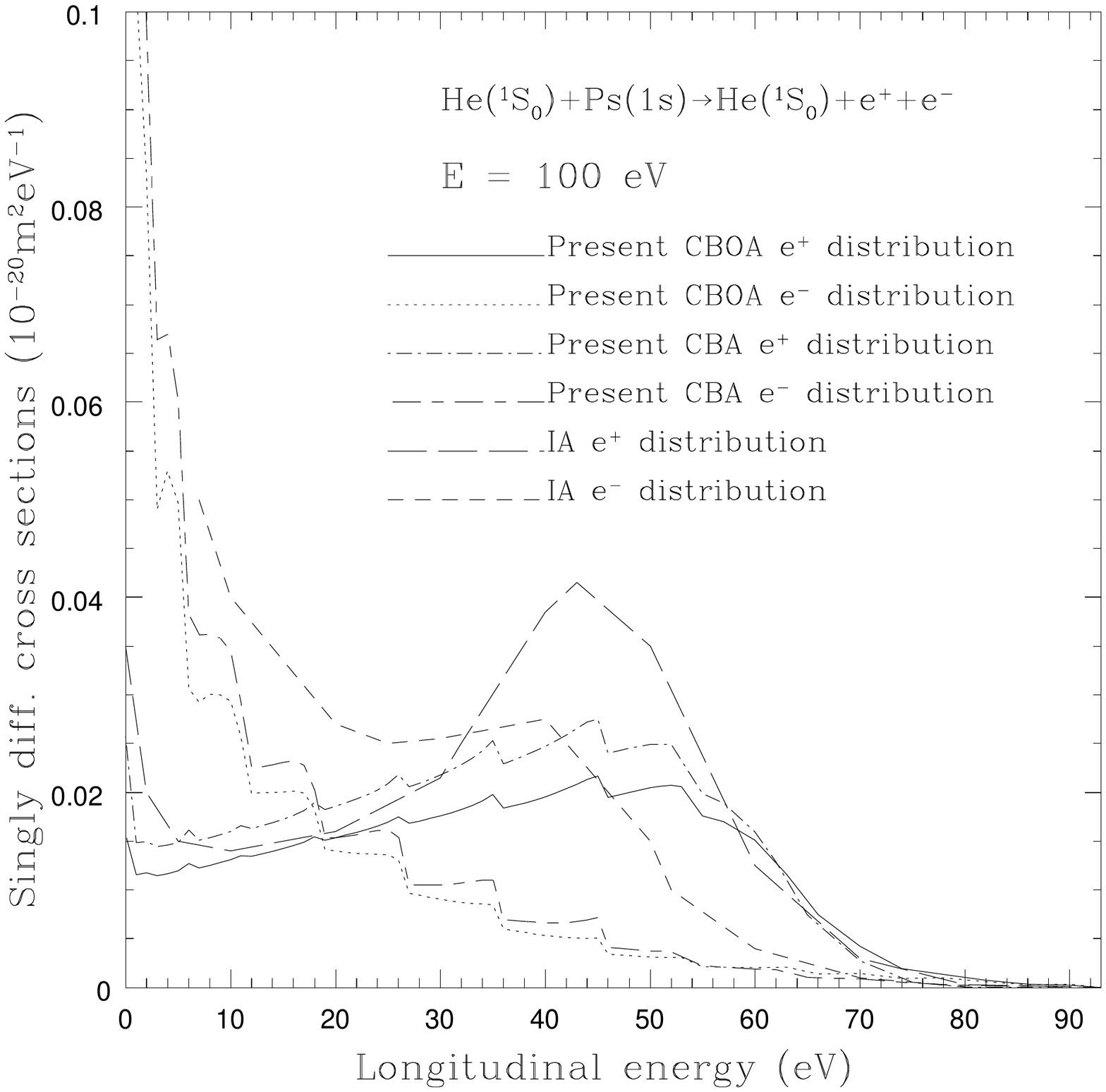}
\caption{}
\end{figure}

\end{document}